\documentclass[runningheads]{llncs}
\usepackage[utf8]{inputenc}
\usepackage[T1]{fontenc}
\usepackage{xcolor}
\usepackage[inline]{enumitem}
\usepackage{booktabs}
\usepackage{amsmath}
\usepackage{amsfonts}
\usepackage{amssymb}
\usepackage{enumitem}
\usepackage{mathbbol}
\usepackage{mathtools}
\usepackage{hyperref}
\usepackage{url}
\usepackage{nicefrac}
\usepackage{microtype}
\usepackage[numbers,sort&compress]{natbib}
\usepackage{multirow}
\usepackage{graphicx}
\usepackage{subcaption}
\usepackage{etaremune}
\usepackage{balance}
\usepackage[font=small]{caption}

\newcommand{\ie}{\emph{i.e.}}

\begin{document}

\title{Revisiting Popularity and Demographic Biases in Recommender Evaluation and Effectiveness}
\titlerunning{Revisiting Popularity and Demographic Biases}

\author{Nicola Neophytou\inst{1} \and
Bhaskar Mitra\inst{2} \and
Catherine Stinson\inst{3}}

\institute{The University of Manchester, Oxford Rd, Manchester M13 9PL, UK 
\email{neophytounicola@gmail.com}
\and
Microsoft, 6795 Rue Marconi, Montréal, Quebec, H2S 3J9, Canada 
\email{bmitra@microsoft.com}\\
\and
School of Computing, 557 Goodwin Hall,
Queen's University,
Kingston ON, 
K7L 2N8, Canada
\email{c.stinson@queensu.ca}}

\maketitle

\begin{abstract}
Recommendation algorithms are susceptible to popularity bias: a
tendency to recommend popular items even when they fail to meet
user needs. A related issue is that the recommendation quality can
vary by demographic groups. Marginalized groups or groups that
are under-represented in the training data may receive less relevant
recommendations from these algorithms compared to others. In a recent study,
\citet{ekstrand2018all} investigate how recommender performance varies according to popularity and demographics, and find statistically significant differences in recommendation utility between binary genders on two datasets, and significant effects based on age on one dataset.  
Here we reproduce those results and extend them with additional analyses.
We find statistically significant differences in recommender performance by both age and gender.
We observe that recommendation utility steadily degrades for older users, and is lower for women than men.
We also find that the utility is higher for users from countries with more representation in the dataset.
In addition, we find that total usage and the popularity of consumed content are strong predictors of recommender performance and also vary significantly across demographic groups.

\end{abstract}

\section{Introduction}
\label{sec:intro}
Recommendation systems and search tools increasingly mediate our access to information online, including news, entertainment, academic resources, and social connections. 
When evaluating the quality of theses results, it is common to report the mean performance over all users.
Majority groups therefore tend to dominate overall statistics when measuring the utility of search and recommendation tools, but utility may also vary across individuals and demographic groups.
Smaller demographic groups, whose needs differ from those of the largest groups, may not be well served by these algorithms that are optimized for mean performance across all users.
If search and recommendation are unfair, in that the utility of search results and recommendations are systematically lower for some demographic groups, members of those groups may be hindered in their decision-making abilities, access to relevant information, and access to opportunities. 

While typical methods of evaluating the effectiveness of search tools and recommender systems do not consider the disparate impact across demographic groups, several recent papers support the concern that these differences in utility do exist.
\citet{mehrotra2017auditing} investigate how the needs of different subgroups of the population are satisfied in the context of search.
In particular, they study the impact on search quality by gender and age and find that both query distribution and result quality vary across these groups.
\citet{ekstrand2018all} perform a similar study in the context of recommender systems, which they investigate through offline top-$n$ evaluation.

In our work, we reproduce the findings by Ekstrand et al., and extend the analysis to incorporate additional user attributes, such as the user's country, usage, and the popularity of the content they consume.
Like them, we find statistically significant differences in recommender utility by age and gender. We further investigate this effect by employing different binning strategies and metrics, and find that, on one dataset, when users are binned by age to achieve roughly equal numbers of users per bin, performance steadily degrades for older users. We also observe recommendation utility on average is higher for men than for women.
In addition, we find the utility is higher for users from countries with more representation in the dataset. To understand how
different demographic attributes impact recommendation quality
relative to each other, we train an Explainable Boosting Machine
(EBM) with user statistics and demographics as features, and
recommender performance as the target variable.
Our results indicate usage and popularity of consumed content are strong predictors of recommender performance.
Both usage and content popularity vary significantly across groups and may provide a partial explanation for the observed differences in recommender utility, though low utility could also partially explain low usage.
In summary, this work studies the following research questions in context of recommender systems:

\newcommand{\RQone}{\begin{itemize}
    \item[\textbf{RQ1}] Does utility vary by demographic group?
\end{itemize}}
\RQone
\newcommand{\RQtwo}{\begin{itemize}
    \item[\textbf{RQ2}] Does utility vary by usage and content popularity?
\end{itemize}}
\RQtwo
\newcommand{\RQthree}{\begin{itemize}
    \item[\textbf{RQ3}] Can usage and popularity explain demographic differences?
\end{itemize}}
\RQthree
\section{Related work}
\label{sec:related}

Recommender systems predict future user-item interactions based on past user-item interactions~\citep{ricci2011introduction}.
Past interactions are often subject to biases---such as selection bias~\citep{marlin2007collaborative}, conformity bias~\citep{krishnan2014methodology, liu2016you}, exposure bias~\citep{liu2020general}, and position bias~\citep{JGPHRG2007, hofmann2014eye, collins2018study}---and the collected data may reflect societal biases towards historically marginalized groups~\citep{karimi2018homophily, stoica2018algorithmic}.
Recommendation algorithms trained on these datasets may further amplify these biases~\citep{zhao2017men, stinson2021algorithms} resulting in homogeneity of recommendations and reduced utility to the user~\citep{chaney:recsys-feedback-loops, hashimoto2018fairness}.
Recommender systems often demonstrate popularity bias~\citep{abdollahpouri2020multi, abdollahpouri2020connection} where popular items are recommended more frequently than warranted by their popularity, and give lower quality recommendations to users with atypical tastes~\citep{burke2002hybrid, ghazanfar2011fulfilling, gras2015users}.
These biases in recommendation raise fairness concerns for all stake-holders~\citep{DBLP:journals/corr/Burke17aa, abdollahpouri2019multi, patro2020fairrec}.
For content producers, unfairness may involve disparate exposure over items of comparable relevance~\citep{diaz2020evaluating, singh2018fairness}.
For consumers of these systems, unfairness may manifest in the form of different recommendation quality across demographic groups~\citep{ekstrand2018all}.
In this work, our focus is on consumer-side fairness, building on prior work by~\citet{ekstrand2018all}.

The fairness concerns in recommendation tasks are not just theoretical questions; they often result in real-world harms.
For example, women may see fewer recommendations for high-paying jobs and career coaching services compared to men~\citep{lambrecht2019algorithmic, datta2015automated}. 
In the context of social networks, previous work ~\citep{stoica2018algorithmic, karimi2018homophily} finds that friend recommender systems can reinforce historical biases by under-recommending minorities.
Unfairness observed on microlending platforms can contribute to certain groups receiving systemically smaller loans, or higher interest rates~\citep{liu2019personalized}.
In ride-hailing platforms, bias can lead to producer-side starvation and loss of income for drivers~\citep{suhr2019two, wang2020faircharge}. Similarly,
\citet{ekstrand2021exploring} find that recommender systems for books disproportionately favor male authors.
The cost to publishers due to under-exposure of their content can be further aggravated by \emph{superstar economics}, common in music and other recommendation scenarios~\citep{rosen1981economics, celma2009music, mehrotra2018towards, ferraro2019music}.
For an overview of fairness and bias in recommender systems, we point the reader to a recent survey by \citet{chen2020bias, ekstrand2021fairness}.
\section{Demographics and Popularity}
Like Ekstrand et al., we begin our analysis with age and binary gender.
For age, in addition to their bucketing scheme, which had unequal age ranges and numbers of users per bucket, we use two additional schemes, such that each age bucket:
\begin{enumerate*}[label=(\roman*)]
    \item is equal in age range, and
    \item includes a roughly equal number of users.
\end{enumerate*}
This analysis with the age attribute is only possible with LASTFM360K data, since ML1M users can only select the age bracket they belong to, as opposed to specifying their exact age in years. This prevents the ability to manipulate age buckets for ML1M. 
We also look at how performance varies by country.
We bucket countries by the number of users in the dataset, and by the country's gross domestic product (GDP)\footnote{https://data.worldbank.org/indicator/NY.GDP.PCAP.CD}, a proxy for socioeconomic status and cultural hegemony.

Users who have interacted more with the recommender system are likely to receive more relevant recommendations.
To analyze how usage influences recommender utility, we bucket users by their number of interactions with items in the collection.
We are also interested in the impact of popularity bias.
The system may do a better job of recommending items to users who typically interact with items that are popular, compared to users with more niche interests.
To investigate how item popularity affects utility, we introduce a novel \emph{pop-index} attribute, defined as the largest value of $p$ such that $p\%$ of items the user has interacted with have also received interactions from $p\%$ of other users.
We take inspiration from the h-index~\citep{hirsch2005index}, used to measure scholarly impact.
We compare recommender utility for groups of users bucketed by pop-index.
\section{Method}
\label{sec:method}

\subsection{Datasets}
\label{sec:analysis-data}
Similar to Ekstrand et al., we conduct our experiments on Last.FM (LFM360K)~\citep{aoscar2010music} and MovieLens (ML1M) data~\citep{DBLP:journals/tiis/HarperK16}.
LFM360K\footnote{\url{http://ocelma.net/MusicRecommendationDataset/lastfm-360K.html}} represents a music recommendation task, and contains $358,868$ users and $292,385$ artists.
For each user-artist pair, the dataset provides the total number of plays.
There are $17,535,605$ user-artist pairs with at least one play in the dataset, which implies that the full user-artist matrix is $99.98\%$ sparse.
Entries in the user-artist matrix were collected using ``user.getTopArtists()'' in the Lastfm API, so include only the top artists for each user, representing a ``playlist'' of their favourite artists.
The number of artists listened to by each user varies across users, with values between one and 166, and a mean of 50.
The dataset also contains user attributes, such as binary\footnote{
We treat gender as a binary class due to the available attributes in the dataset.
We do not intend to suggest that gender identities are binary.}
gender ($67$\% male, $24$\% female, $9$\% missing), age ($20$\% missing), and country (none missing).

Our second dataset ML1M\footnote{\url{https://grouplens.org/datasets/movielens/1m/}} represents a movie recommendation task.
ML1M contains $3,952$ movies and $6,040$ users who joined MovieLens in 2000.
Each user-movie pair has an associated 5-point rating assigned by the user.
The dataset contains $1,000,209$ ratings, corresponding to a $95.81\%$ sparse user-movie matrix.
Each user has rated at least 20 movies.
The dataset also includes a binary gender, age, and occupation for each user.
For the ML1M data set, users can only specify that they belong to a pre-set age bracket, as opposed to specifying exactly how old they are in years. The choice of age brackets they can choose from are displayed on the x-axis of Figure 1g.



\subsection{Model}
\label{sec:analysis-model}
We use an alternating Least Squares model for implicit feedback data
~\citep{hu2008collaborative}, as implemented in the Implicit\footnote{\url{https://github.com/benfred/implicit}} code repository.
We use the default hyperparameters as recommended by Implicit, by setting factors to $50$ and the regularization constant to $0.01$.
We train the model for $30$ iterations in all experiments.
The Implicit code performs some data cleanup - as described here\footnote{\url{https://github.com/benfred/bens-blog-code/blob/master/distance-metrics/musicdata.py\#L39}}- to deal with malformed entries in the data files.
All statistics reported in Section~\ref{sec:results} are computed after this cleanup.

\subsection{Experiment protocol}
\label{sec:analysis-experiment}
We conduct our experiments under a five-fold cross-validation setting.
For LFM360K, each test partition contains $5,000$ randomly sampled users. For ML1M we partition the whole set of $6,040$ users into five splits containing $1,208$ users, for each iteration of cross-validation. 
For both datasets, we hold out $20\%$ of the items each user has interacted with to use as test data.
All other users and the rest of the test users' items are used for model training in each iteration.
To avoid the cold-start problem, we remove users who listened to 40 or fewer artists in the LFM360K dataset--roughly $10\%$ of users.
The ML1M dataset only includes users who have rated over 20 or more movies, so none are removed.
For evaluation, we generate $1,000$ recommendations per user, and measure the results using NDCG, MRR, and RBP metrics.
To verify if differences in utility are significant across demographics, we perform Kruskall-Wallis significance tests on mean NDCG values between the demographic groups. For attributes which contain an N/A group, where the information on this attribute is not provided by the user, the N/A group is omitted from Kruskall-Wallis testing. This ensures we are only comparing groups of users who provided information on this attribute.
We also run Bonferroni correction for multiple testing.

To understand the relative impact of user attributes on system performance, we train an Explainable Boosting Machine (EBM) model, as implemented in the InterpretML framework~\citep{nori2019interpretml}, to predict the mean NDCG for each user as a dependent variable.
We represent each user by a combination of the following features:
\begin{enumerate*}[label=(\roman*)]
    \item Age,
    \item Gender,
    \item Country, ordered by prevalence in the dataset and bucketed (LFM360K only),
    \item Country, ordered by GDP and bucketed (LFM360K only),
    \item Usage (\ie, total number of listens for LFM360K and total number of movies rated for ML1M),
    \item Pop-index, and finally
    \item The last digit of the user ID.
\end{enumerate*}
The last digit of the user ID serves as a control feature which should have no effect on performance on either dataset.
We run the EBM model once individually for each feature group, and once with all features included for cross feature-group comparison.
\section{Results}
\label{sec:results}

Using the datasets and methods described above, we reproduce the main results from Ekstrand et al., and inquire in more detail how the quality of recommendation varies by age, gender, and country, using varied binning strategies and metrics.
In addition, we study the impact of usage and item popularity on utility, and how they interplay with the other demographic variables.

\begin{figure}[t]
\center
\begin{subfigure}{1.0\textwidth}
    \includegraphics[width=\textwidth]{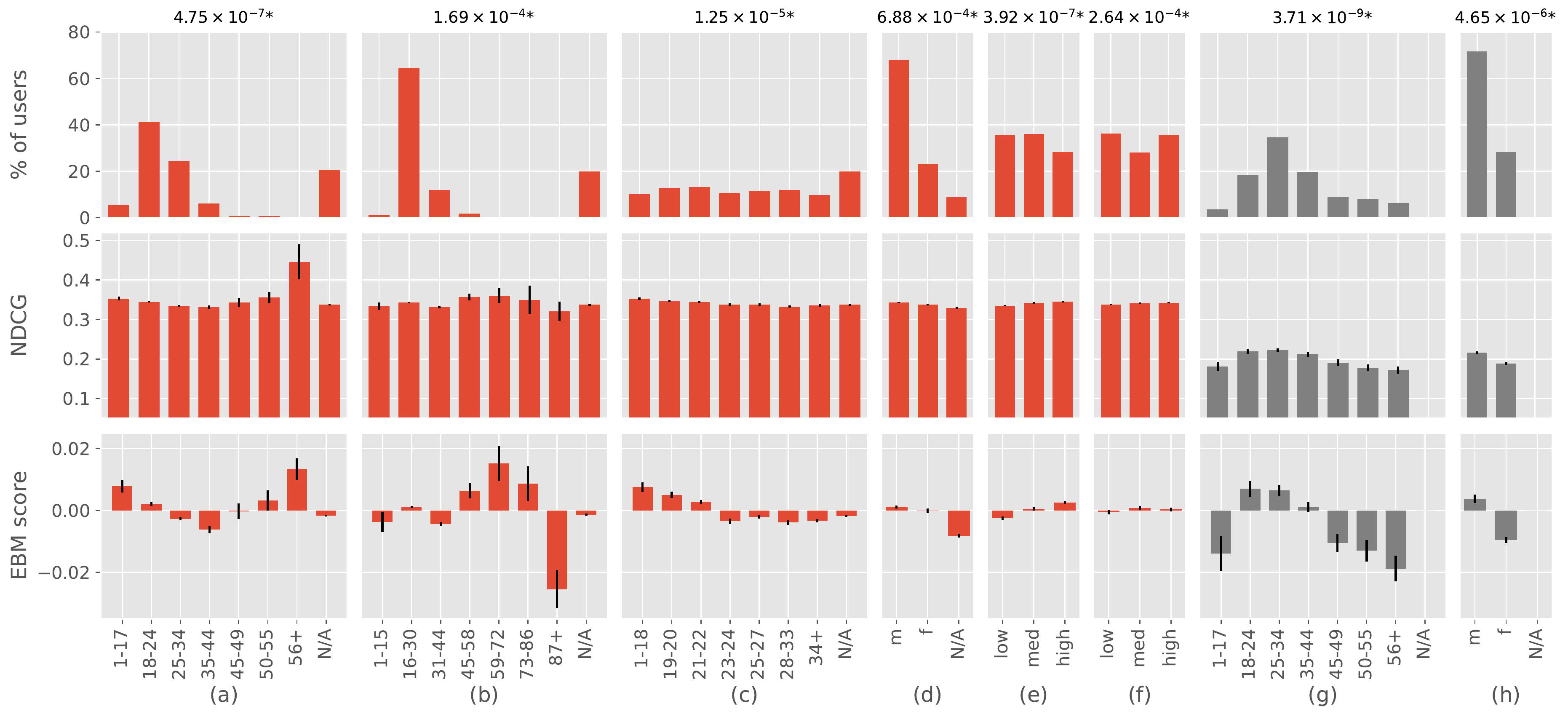}
\end{subfigure}

\caption{ Comparison of binning strategies, metrics, and datasets on recommender utility by demographic variables. Red plots represent the LFM360K dataset and grey represent ML1M. For age, we consider the original bucketing scheme from Ekstrand et al. (a and g), and buckets by equal range (b) and equal number of users (c). (d) and (h) represent gender for LFM360K and ML1M, respectively. (e) and (f) represent country ordered by number of users and by GDP for LFM360K. P-values from Kruskal-Wallis significance tests on NDCG are reported above each column.}
\label{fig:analysis-demographic}
\end{figure}

\RQone
Figure \ref{fig:analysis-demographic} shows the distribution of users, recommender utility ( mean NDCG), and the EBM scores corresponding to different demographic variables.
Figure 1a-1f corresponds to the LFM360K dataset (in red). Column (a) divides users into age groups according to the age range
buckets Ekstrand et al. used, replicating their results. Column (b) divides users into age buckets of uniform range (15 years). Column
(c) organizes users into age buckets such that the number of users in
each bucket is comparable. Figure
1g-1h presents the results for the ML1M dataset (in grey), where the age buckets again correspond to those used in Ekstrand et al., replicating their results.
For each column, we run the Kruskall-Wallis significance test and on all metrics. P-values for mean NDCG are reported above each column.

\subsection{Impact on age}
Ekstrand et al. find significant differences in recommender utility across different user age brackets according to the Kruskal-Wallis significance test.
Our analysis confirms these findings on both datasets, as we also report significant differences based on Kruskal-Wallis significance test ($p < 0.01$) across the same age brackets 
(Figure 1a and 1g). 
We also find significant differences when we try alternative binning strategies on LFM360K, corresponding to bins with equal age range
(Figure 1b) and bins with equal number of users (Figure 1c).
While we only report p-values corresponding to the NDCG metric for recommendation utility, we have verified the differences are also statistically significant for MRR and RBP, except for MRR for ML1M.

The first row shows on both datasets that the age distribution is skewed towards young adults, more so for LFM360K than ML1M.
Because the age buckets were irregular, we show the results with buckets of uniform range
(Figure 1b).
We also posit that a skewed distribution of users across age buckets may make it difficult to detect differences in utility across ages, because some age buckets contain very few users.
Therefore, we additionally try buckets containing approximately equal numbers of users
(Figure 1c).
When the number of users in each bucket are comparable, we find a gradual downward trend in recommender utility, as age increases.
This effect was not visible in Ekstrand et al.
We also observe a similar downward trend on ML1M as seen in
Figure 1g.
This trend is further confirmed by the EBM scores in
Figures 1c and 1g where younger ages correspond to higher EBM scores when the number of users in each bucket are approximately equal. 

\subsection{Impact on gender}
Both LFM360K (Figure 1d) and ML1M (Figure 1h) datasets contain many more male than female users.
As in Ekstrand et al., we observe statistically significant differences in utility by gender based on Kruskal-Wallis significance test ($p < 0.01$), with better recommendation utility for male than female users. This is observed in both datasets, except for MRR and RBP for LFM360K, and MRR for ML1M. Given the unbalanced user distribution across genders in these datasets, this can either be the result of a popularity bias, or a demographic bias.
We revisit this question later in this section in the context of RQ3.

\subsection{Impact on country}
An additional demographic variable available in the LFM360K dataset, but not in ML1M, is users' country of residence.
Ekstrand et al. did not analyze whether there is evidence of recommender utility differences by country, but we perform this analysis here.
We group the countries in two ways.
First, according to its representation in the dataset---\ie, based on the number of users from that country, into low, medium, and high buckets ---and second, by GDP, again into low, medium, and high buckets.
Figures 1e and 1f show the results corresponding to the two analyses. Low GDP is used here as a proxy for social marginalization.

We find statistically significant differences by country on both measures, except for MRR and RBP for GDP.
The model has higher recommender utility for users from countries with more representation in the dataset. The same trend is not observed, however, when countries are ordered by GDP.

As expected, there are no statistically significant differences found on any metric between users grouped by the last digit of their user ID, the control feature, across both data sets.

\begin{figure}[t]
\center
\begin{subfigure}{0.9\textwidth}
    \includegraphics[width=\textwidth]{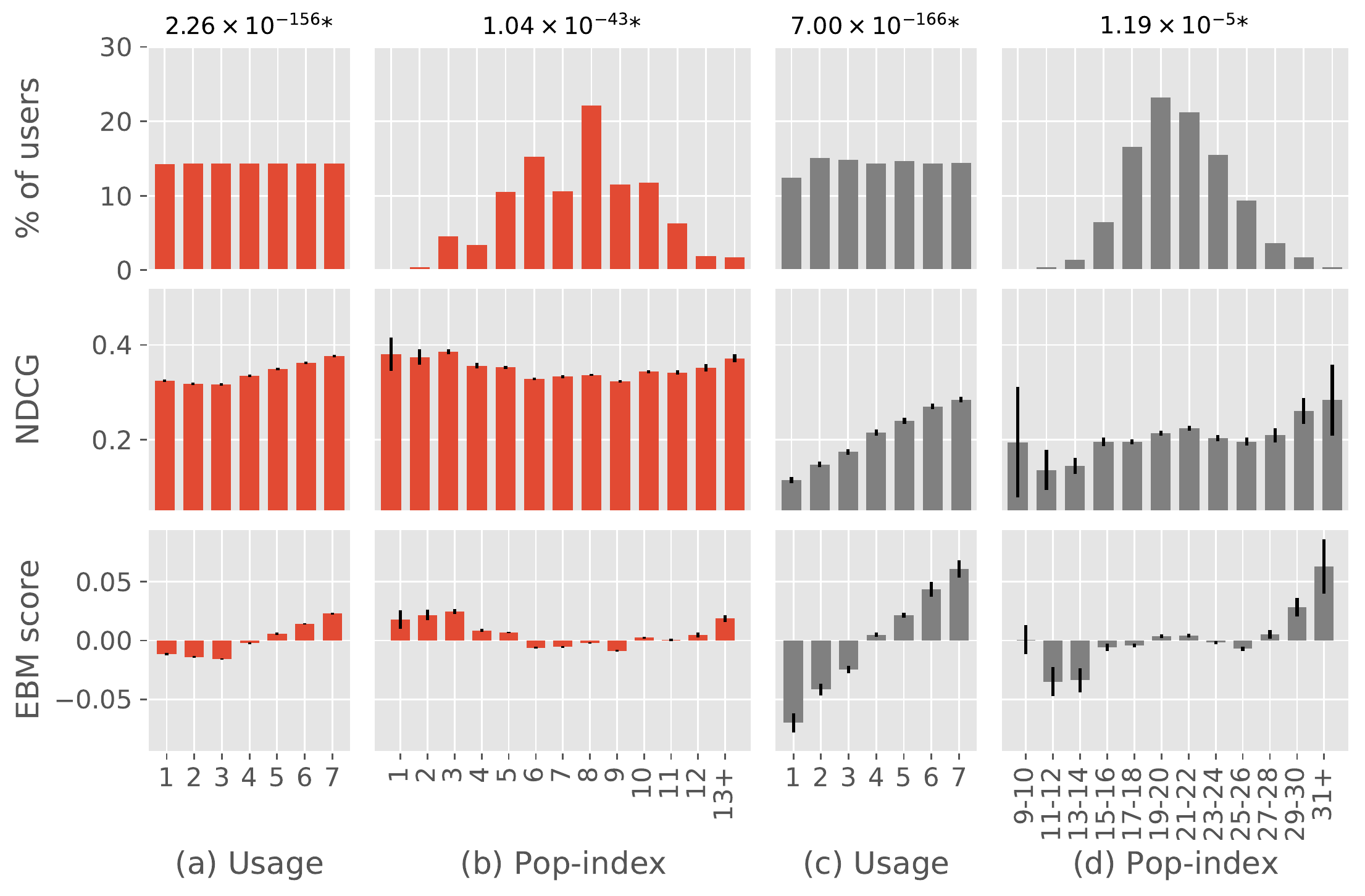}
\end{subfigure}

\caption{Recommendation utility by usage and content popularity. Red plots represent the LFM360K dataset, grey plots represent ML1M. p-values from Kruskal-Wallis significance tests on NDCG are reported above each column.}
\label{fig:analysis-usage-and-pop}
\end{figure}

\filbreak
\RQtwo
It is not obvious when to attribute utility differences across groups of users to popularity bias, rather than bias specifically affecting demographic groups, because marginalized groups are often also less represented in training datasets.
To explore this issue, we first investigate how recommender utility is affected by two measures of popularity: usage and pop-index.
For a given user, high usage implies more representation in the data, while a higher pop-index corresponds to affinity towards items that are popular with other users in the dataset.
In Figure \ref{fig:analysis-usage-and-pop} we compare both these measures on the LFM360K and ML1M datasets.
For both datasets there is a trend toward greater NDCG as usage increases. 
The EBM analysis shows the same trend, where low usage corresponds to a negative effect on the EBM score, and high usage corresponds to a positive effect.
We also investigate popularity in the sense of how popular items preferred by a user are among the user population as a whole.
Our hypothesis is that users whose playlists contain more popular items will likely have greater recommendation utility.
On ML1M (Figure 2d), we observe a trend which supports our hypothesis.
However, on LFM360K
(Figure 2b), we observe a U-shaped trend, with higher utility associated with both groups of users with maintstream and unique tastes.
We suspect differences in observations on the two datasets may be partially explained by the semantics of user interactions in the two cases.
In LFM360K, the user interacts with an artist by listening to them, and they can listen to the same artist multiple times.
So, for users with more distinctive tastes, the recommender algorithm may still achieve reasonable performance by recommending items the user interacted with before.
In contrast, in ML1M the user interacts with the item by providing a rating and therefore the recommender must suggest new items the user has not interacted with before, which is a more difficult challenge, specifically when the user has a distinctive taste.

\begin{figure}[t]
    \begin{subfigure}{0.5\textwidth}
        \includegraphics[width=\textwidth]{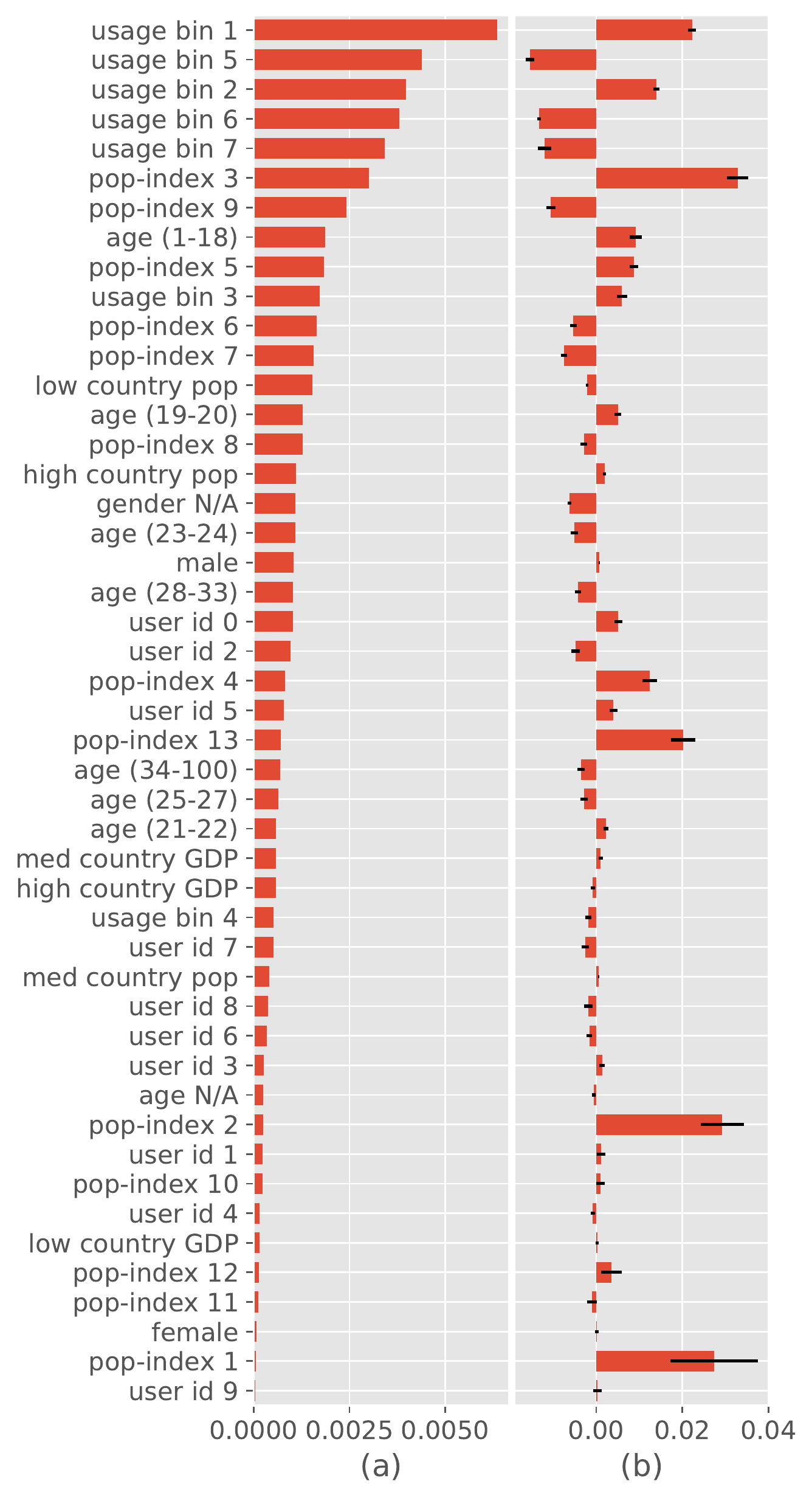}
    \end{subfigure}
    \hfill
    \begin{subfigure}{0.5\textwidth}
        \includegraphics[width=\textwidth]{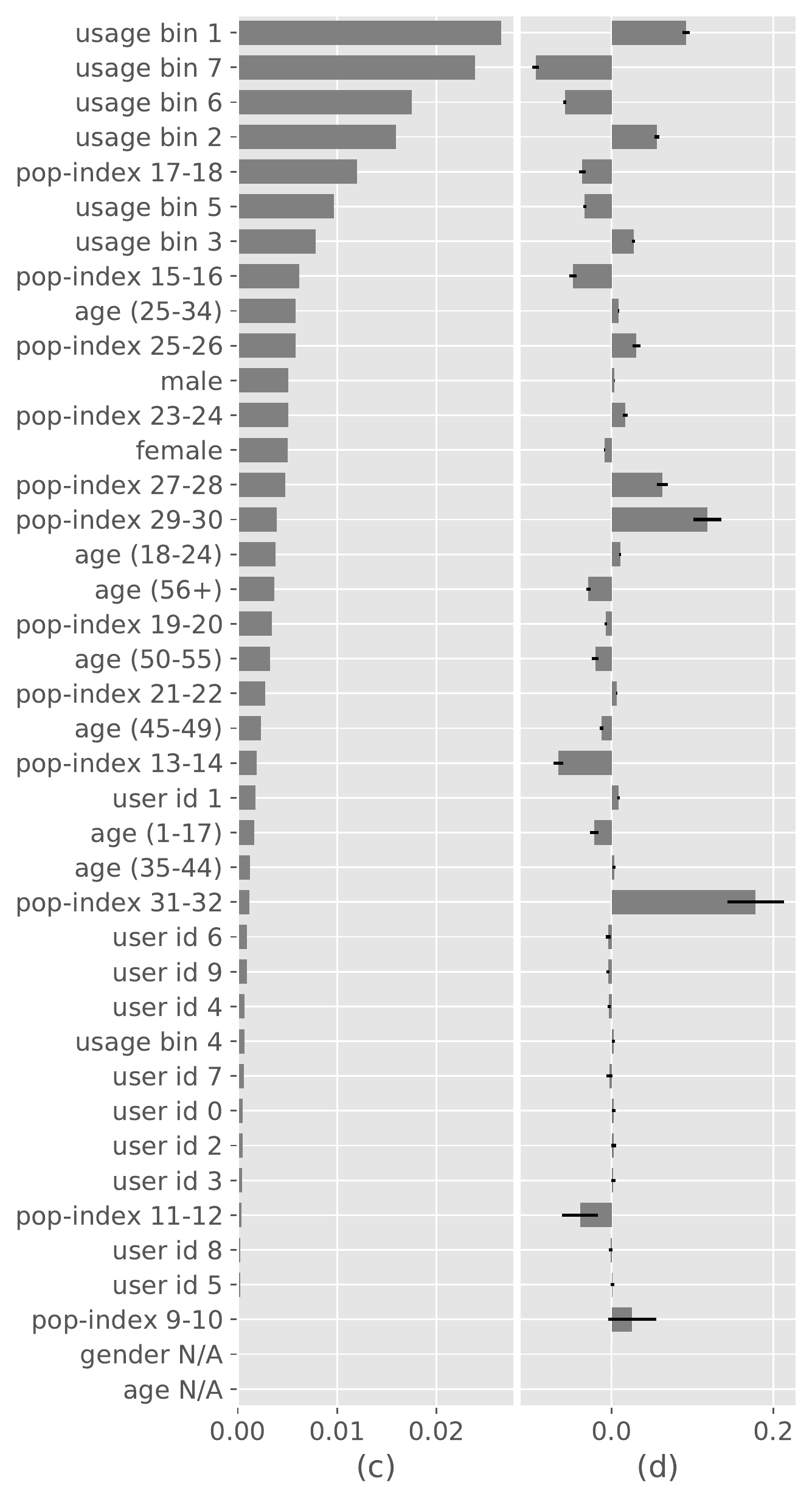}
    \end{subfigure}
\caption{Ranked features and their scores from the EBM analysis. In (a) and (c) equal numbers of users are sampled for each factor. In (b) and (d) the full database is used.}
\label{fig:EBM-plot}
\end{figure}

\RQthree
One of our goals is to better understand the relative importance of different demographic and popularity features to explain the differences in mean recommender utility amongst users.
Towards that goal, we train an EBM model to predict mean recommender utility based on these user attributes.
Figure~\ref{fig:EBM-plot} shows that on both datasets (LFM360K and ML1M) the usage features emerge as the most predictive, followed by pop-index.
Among the demographic attributes, some of the age-related features are ranked highest on both datasets.
On LFM360K, age is followed by country (ordered by number of users) and gender as the next most predictive user attributes.
In the absence of country information, on the ML1M dataset we observe gender to be high in the feature ranking after age.
The high feature importance for usage and pop-index provides evidence than some of the demographic differences may be explained by representation in the data.
This is not to argue that the recommender system under study is fair to different demographics of users.
Disparity of utility across demographics may directly influence user retention~\citep{ekstrand2018all} and usage.
This creates a vicious cycle where a small difference in utility across user groups may be further amplified by subsequent disparity in system adoption and usage across demographics, leading to even bigger disparities in utility.
Table~\ref{tbl:results} shows how usage and pop-index are distributed across demographic groups, further demonstrating how they may correlate with historical marginalization.

\begin{table}[t]
    \centering
    \caption{Percentage of users in different usage and pop-index buckets for each demographic groups in LFM360K.
    For younger users and men a higher proportion of the population correspond to higher usage buckets.
    The trend for pop-index is less clear.}
    \begin{tabular}{cr|rrrrrrrrr|rrr}
    \hline
    \hline
    & & \multicolumn{8}{c}{\textbf{Age} (bucketed by equal number of users)} & & \multicolumn{3}{c}{\textbf{Gender}} \\
    & & \textbf{1-18} & \textbf{19-20} & \textbf{21-22} & \textbf{23-24} & \textbf{25-27} & \textbf{28-33} & \textbf{34+} & \textbf{N/A} & & \textbf{m} & \textbf{f} & \textbf{N/A} \\
    \hline
    \textbf{Usage} & & & & & & & & & & & & & \\
    $1$ & & $11\%$ & $ 8\%$ & $ 9\%$ & $11\%$ & $12\%$ & $15\%$ & $24\%$ & $21\%$ & & $13\%$ & $16\%$ & $20\%$ \\
    $2$ & & $13\%$ & $11\%$ & $12\%$ & $13\%$ & $13\%$ & $17\%$ & $17\%$ & $17\%$ & & $14\%$ & $15\%$ & $15\%$ \\
    $3$ & & $14\%$ & $15\%$ & $13\%$ & $13\%$ & $15\%$ & $15\%$ & $15\%$ & $14\%$ & & $14\%$ & $16\%$ & $15\%$ \\
    $4$ & & $16\%$ & $15\%$ & $15\%$ & $14\%$ & $14\%$ & $14\%$ & $13\%$ & $14\%$ & & $14\%$ & $15\%$ & $13\%$ \\
    $5$ & & $15\%$ & $16\%$ & $15\%$ & $16\%$ & $16\%$ & $14\%$ & $11\%$ & $13\%$ & & $14\%$ & $14\%$ & $13\%$ \\
    $6$ & & $15\%$ & $17\%$ & $18\%$ & $17\%$ & $14\%$ & $13\%$ & $11\%$ & $11\%$ & & $15\%$ & $13\%$ & $12\%$ \\
    $7$ & & $17\%$ & $18\%$ & $18\%$ & $16\%$ & $15\%$ & $12\%$ & $ 9\%$ & $10\%$ & & $16\%$ & $11\%$ & $11\%$ \\
    \hline
    \textbf{Pop-index} & & & & & & & & & & & & & \\
    $ 1$ & & $ 0\%$ & $ 0\%$ & $ 0\%$ & $ 0\%$ & $ 0\%$ & $ 0\%$ & $ 0\%$ & $ 0\%$ & & $ 0\%$ & $ 0\%$ & $ 0\%$ \\
    $ 2$ & & $ 0\%$ & $ 0\%$ & $ 0\%$ & $ 0\%$ & $ 0\%$ & $ 0\%$ & $ 1\%$ & $ 1\%$ & & $ 0\%$ & $ 0\%$ & $ 0\%$ \\
    $ 3$ & & $ 3\%$ & $ 3\%$ & $ 3\%$ & $ 4\%$ & $ 4\%$ & $ 5\%$ & $ 8\%$ & $ 6\%$ & & $ 5\%$ & $ 4\%$ & $ 5\%$ \\
    $ 4$ & & $ 2\%$ & $ 2\%$ & $ 3\%$ & $ 3\%$ & $ 3\%$ & $ 3\%$ & $ 5\%$ & $ 4\%$ & & $ 4\%$ & $ 2\%$ & $ 4\%$ \\
    $ 5$ & & $ 7\%$ & $ 9\%$ & $ 9\%$ & $10\%$ & $11\%$ & $12\%$ & $15\%$ & $11\%$ & & $11\%$ & $ 8\%$ & $11\%$ \\
    $ 6$ & & $15\%$ & $15\%$ & $14\%$ & $16\%$ & $14\%$ & $15\%$ & $15\%$ & $17\%$ & & $15\%$ & $14\%$ & $17\%$ \\
    $ 7$ & & $10\%$ & $10\%$ & $10\%$ & $10\%$ & $11\%$ & $10\%$ & $12\%$ & $11\%$ & & $11\%$ & $10\%$ & $11\%$ \\
    $ 8$ & & $23\%$ & $23\%$ & $23\%$ & $23\%$ & $24\%$ & $22\%$ & $19\%$ & $20\%$ & & $21\%$ & $25\%$ & $21\%$ \\
    $ 9$ & & $14\%$ & $13\%$ & $12\%$ & $12\%$ & $11\%$ & $10\%$ & $ 9\%$ & $11\%$ & & $11\%$ & $13\%$ & $12\%$ \\
    $10$ & & $14\%$ & $12\%$ & $13\%$ & $12\%$ & $12\%$ & $12\%$ & $10\%$ & $10\%$ & & $12\%$ & $13\%$ & $ 8\%$ \\
    $11$ & & $ 7\%$ & $ 8\%$ & $ 7\%$ & $ 6\%$ & $ 6\%$ & $ 6\%$ & $ 4\%$ & $ 5\%$ & & $ 6\%$ & $ 7\%$ & $ 6\%$ \\
    $12$ & & $ 2\%$ & $ 3\%$ & $ 2\%$ & $ 2\%$ & $ 2\%$ & $ 2\%$ & $ 1\%$ & $ 2\%$ & & $ 2\%$ & $ 2\%$ & $ 2\%$ \\
    $13+$ & & $ 2\%$ & $ 2\%$ & $ 2\%$ & $ 2\%$ & $ 1\%$ & $ 2\%$ & $ 1\%$ & $ 1\%$ & & $ 2\%$ & $ 2\%$ & $ 1\%$ \\
    \hline
    \end{tabular}
    \label{tbl:results}
\end{table}
\section{Discussion and conclusion}
\label{sec:conclusion}

We confirmed that recommender systems are prone to unfairness across the demographic attributes available in the datasets used here. To explore this question more thoroughly, one would need access to more detailed demographic data, and the ability to observe temporal dynamics of how recommendations affect usage and usage affects recommendations. In order to answer questions like what caused the U-shaped pattern we found in recommender utility by usage, we would need the ability to intervene on recommendations in real time. 


\citet{mehrotra2017auditing} point out that users for whom a search engine is least satisfactory can paradoxically end up having the highest measured utility. They found when utility is bad enough to make a user stop using the service for everyday needs, they still use the search engine for very easy queries that they assume even a poor search engine could get right. Such searches end up being successful, resulting in artificially high utility scores. User attrition is an issue we cannot track given the datasets used here. It may be that users who have the highest usage are a self-selecting group for whom recommenders happen to work well. 

For both datasets there is a trend toward greater utility as usage increases. This is unsurprising, given that users with higher usage will provide more labels, with which the recommender can build a more accurate model of user preferences. 
One anomalous effect we observed is in the LastFM dataset; users with least usage have higher utility recommendations than users with slightly more usage. This could be evidence of the same effect as observed by \citet{mehrotra2017auditing}. If LastFM gives poor recommendations for a given user, that user might stop using it for everyday music streaming, but still use it when they are looking for something very mainstream. Another possibility is since LastFM users input a few artists they like when setting up their accounts, early listens will be dominated by artists which the user identified as being among their favourites, rather than recommendations provided by the model. Utility may therefore be artificially high during early use.



The social harms that can result from unfair recommendation go well beyond some people choosing not to use a tool that others find fun and convenient. Recommendation algorithms are increasingly being used to make major life decisions, like mortgage lending, job searching, connecting with community, and basic access to information. The body of work we are adding to here demonstrates that fair recommendation is a problem requiring serious attention.

\bibliographystyle{splncs04nat}
\bibliography{bibtex}
\end{document}